\documentclass[12pt]{article}
\usepackage[english]{babel}
\usepackage{graphicx,epsfig}
\makeindex \textwidth 170mm \textheight 220mm \topmargin -5mm
\newcommand{\be}{\begin{equation}}
\newcommand{\ee}{\end{equation}}
\newcommand{\bea}{\begin{eqnarray}}
\newcommand{\eea}{\end{eqnarray}}

\def\derp#1#2{\rp{\partial{#1}}{\partial{#2}}}
\def\rfr#1{eq.(\ref{#1})}

\def\Rfr#1{Eq.(\ref{#1})}

\def\bb{\bibitem}
\def\eqi{\begin{equation}}
\def\eqf{\end{equation}}
\def\eqia{\begin{eqnarray}}
\def\eqfa{\end{eqnarray}}

\def\rp#1#2{{#1\over#2}}

\def\lb#1{\label{#1}}

\begin{document}
\begin{titlepage}
\begin{flushright}
\today\\
BARI-TH/00\\
\end{flushright}
\vspace{.5cm}
\begin{center}
{\LARGE Is it possible to improve the present LAGEOS--LAGEOS II
Lense--Thirring experiment?} \vspace{1.0cm}
\quad\\
{Lorenzo Iorio$^{\dag}$\\
\vspace{0.5cm}
\quad\\
{\dag}Dipartimento di Fisica dell' Universit{\`{a}} di Bari, via
Amendola 173, 70126, Bari, Italy}\\ \vspace{1.0cm}

{\bf Abstract\\}
\end{center}

{\noindent \footnotesize The Lense-Thirring effect is currently
being measured by means of a combination of the orbital residuals
of the nodes of LAGEOS and LAGEOS II and the perigee of LAGEOS II.
The claimed total error should be of the order of $20\%$. The most
insidious systematic error is due to the mismodelled even zonal
harmonics of the geopotential and amounts to $12.9\%$, according
to the full covariance matrix of the EGM96 model up to degree
$l=20$. The role and the importance of the LAGEOS--LAGEOS II
Lense--Thirring experiment is investigated. Using other suitable
combinations with orbital elements of the other existing
laser-ranged satellites does not yield significative improvements
except, perhaps, for one combination including the nodes of
LAGEOS, LAGEOS II and Ajisai and the perigee of LAGEOS II. The
related systematic error due to the mismodelled even zonal part of
the geopotential reduces to almost 10.7$\%$, according to the full
covariance matrix of EGM96 up to degree $l=20$. }
\end{titlepage} \newpage \pagestyle{myheadings} \setcounter{page}{1}
\vspace{0.2cm} \baselineskip 14pt

\setcounter{footnote}{0}
\setlength{\baselineskip}{1.5\baselineskip}
\renewcommand{\theequation}{\mbox{$\arabic{equation}$}}
\noindent

\section{Introduction}
%A physical theory can be considered worthy of trust when there
%exists a large collection of empirically checked independent
%consequences of it. The largest is the amount of independent
%predictions of the theory which can be tested, the highest is the
%degree of reliability and confidence gained by the theory.
%Moreover, if we look for more and more accurate tests of the
%theory the possibilities of discovering some errors in it or,
%however, new problems increase greatly. This is just the soul of
%the scientific enterprise.
In regard to General Relativity (GR),
%and other
%metric theories of gravitation parameterized in the context of the
%PPN framework,
the experience shows that designing and effectively performing
experiments aimed to test it is very difficult and often
expensive. This is particularly true for its linearized
weak--field and slow--motion approximation which is adequate in
the Solar System. The relativistic effects quite often fall below
or lie just at the edge of the experimental sensitivity of various
techniques which could be used. Moreover, there are lots of biases
and aliasing effects induced by a host of classical phenomena
which pose severe constraints to the obtainable accuracy in
space-based experiments. In many cases the systematic errors are
larger than the standard statistical errors resulting from the
least-squares procedures with which the relativistic effects are
commonly extracted from the data.

%Regarding the PPN parameters, $\beta$ and $\gamma$ are those
%related to the "classical" tests of GR. At present, they are known
%at a precision level of the order of $10^{-3}$. They have been
%obtained from the measurement of some post-Newtonian effects
%parameterized in terms of them. However, there are other
%post-Newtonian features, obtainable in terms of $\beta$ and/or
%$\gamma$, which have not yet been directly measured.
%According to
%some authors, it would not be particularly interesting to try to
%measure them because of the relatively high accuracy already
%existing in the values of $\beta$ and $\gamma$ obtained from other
%tests. We are referring, e.g., to the Lense-Thirring drag of
%inertial frames and the current LAGEOS experiment aimed to its
%detection.
So, it is of the utmost importance
\begin{itemize}
\item To design experiments capable of effectively testing some
post-Newtonian effects not yet directly checked
%independently of
%the already accepted values for the PPN parameters from other
%tests,
at relatively low expense of time and money
\item To assess as more reliably and clearly as possible the error
budget so to single out the various aliasing competing forces and
to evaluate the impact of their systematic errors \item To
investigate more thoroughly and reanalyze the role played by the
already existing or proposed experiments in order to correctly
evaluate their importance
\end{itemize}
%With respect to these goals the technique of the Satellite Laser
%Ranging plays a key role.
%By analyzing the laser data of the
%orbits of some existing or proposed cannonball passive geodetic
%satellites it has been possible to measure the gravitomagnetic
%Lense-Thirring drag of the orbits of LAGEOS and LAGEOS II due to
%the proper rotation of the Earth at 20$\%$ level. Moreover, it
%should be possible to measure by few years the gravitoelectric
%advance of the pericenter $\omega$ of a test mass due to the
%Schwarzschild part of the metric due to a non-rotating central
%body at a level of $10^{-3}--10^{-4}$. The use of laser--ranged
%satellites should be the natural candidate to measure the
%gravitomagnetic clock effect of a couple of counter-rotating test
%masses as well. Indeed, it has been recently shown that, at least
%regarding the non-gravitational perturbations, the present
%technology of SLR satellites would allow to meet the stringent
%accuracy levels of $10^{-2}$ cm and $10^{-2}$ mas per revolution
%in the knowledge of the satellites'orbits.
%-----------------------------------------------------------
\subsection{The importance of the present Lense-Thirring LAGEOS
experiment} Nowadays, a first attempt to measure the
Lense-Thirring drag of the orbit of a test body [{\it Ciufolini
and Wheeler}, 1995] in the gravitational field of the Earth, by
using the following linear combination of orbital residuals of the
rates of the node $\delta\dot\Omega$ of LAGEOS and LAGEOS II and
the perigee $\delta\dot\omega$ of LAGEOS II [{\it Ciufolini},
1996]\eqi\delta\dot\Omega^{\rm I}+c_1\delta\dot\Omega^{\rm
II}+c_2\delta\dot\omega^{\rm II}\sim \mu_{\rm LT}60.2\lb{uno},\eqf
has been reported in [{\it Ciufolini et al.}, 1998; {\it
Ciufolini}, 2002]. In \rfr{uno} $c_1\sim 0.295$, $c_2\sim-0.35$
are coefficients suitably designed in order to cancel out every
mismodelled contributions of the first two even ($l=2,4$) zonal
($m=0$) harmonics $J_2$ and $J_4$ of the multipolar expansion of
the non-spherical Earth's gravitational field and are built up
with the orbital parameters of LAGEOS and LAGEOS II (see section 3
for further details). Finally, $\mu_{\rm LT}$
%$=\rp{1+\gamma}{2}$
is the parameter in terms of which the Lense-Thirring drag is
expressed: it is 1 in GR and 0 in Newtonian mechanics. GR predicts
for \rfr{uno} a linear trend with a slope of 60.2 milliarcseconds
per year (mas/y). \Rfr{uno} would allow to obtain a claimed total
experimental error of the order of $20\%$ [{\it Ciufolini et al.},
1998; {\it Ciufolini}, 2002]. The important GP--B mission [{\it
Everitt et al.,} 2001], which is aimed to the detection of a
complementary gravitomagnetic effect on freely falling gyroscopes
and which is scheduled to fly at the beginning of 2003, would
reach a claimed accuracy level of 1$\%$. It is worth noting that
the strategy of the combined residuals of LAGEOS and LAGEOS II
could allow to constraining strongly the hypothesis of a
Yukawa--like fifth force [{\it Iorio}, 2002a] and to measure in
the field of the Earth the gravitoelectric relativistic perigee
advance [{\it Iorio et al.}, 2002a].

What is the real importance of the observable of \rfr{uno}? Is it
the only possibility we have at present in order to measure the
Lense-Thirring drag of the orbit of a test body in the terrestrial
space environment [{\it Casotto et al.}, 1990]? Would it be
possible to adopt alternative combinations involving orbital
elements of other SLR satellites? We will try to answer these
questions in the following.

In Table 1 we quote the orbital parameters of the existing SLR
geodetic satellites and of the proposed LARES. In it $a,\ e,\ i$
and $n$ are the semimajor axis, the eccentricity, the inclination
and the  Keplerian mean motion, respectively.

The paper is organized as follows. In section 2 we analyze the
systematic errors affecting such measurement. In section 3 we look
for alternative combinations including the orbital residuals of
other existing SLR satellites. Section 4 is devoted to the
conclusions. {\small{
\begin{table} \caption{Orbital parameters of the
existing spherical passive geodetic laser-ranged satellites and of
LARES. Aj=Ajisai, Stl=Stella, Str=Starlette, WS=WESTPAC--1,
E1=ETALON--1, E2=ETALON--2, L1=LAGEOS, L2=LAGEOS II, LR=LARES. $a$
is in km, $i$ in deg and $n$ in s$^{-1}$. } \label{paras}
\begin{center}
\begin{tabular}{llllllllll}
\hline   & { {Aj}} & {Stl} & {Str} & {WS} & {E1} & {E2} & L1 & L2 & LR\\
\hline
$a$  & 7,870 & 7,193 & 7,331 & 7,213 & 25,498 & 25,498 & 12,270 & 12,163 & 12,270\\
$e$ & 0.001 & 0 & 0.0204 & 0 & 0.00061 & 0.00066 & 0.0045 & 0.014 & 0.04\\
$i$  & 50 & 98.6 & 49.8 & 98 & 64.9 & 65.5 & 110 & 52.65 & 70\\
$n$  & 0.0009 & 0.001 & 0.001 & 0.001 & 0.00015 & 0.00015 &
0.00046 & 0.00047 &
0.00046\\
\hline
\end{tabular}
\end{center}
\end{table}
}}
%------------------------------------------------------------------------------------
\section{The systematic errors} The systematic errors affecting
such measurement are of two types\begin{itemize}
\item Long-periodic aliasing harmonics of gravitational (solid
Earth and ocean tides [{\it Iorio}, 2001; {\it Iorio and Pavlis,}
2001; {\it Pavlis and Iorio}, 2002]) and non-gravitational (direct
solar radiation pressure, Earth albedo, direct Earth IR radiation
pressure, thermal effects [{\it Lucchesi}, 2001; 2002]) origin. If
their periods $P$ are shorter than the observational time span
$T_{\rm obs}$ they can be viewed as empirically fitted quantities
and can be removed from the signal. If $P=kT_{\rm obs}, k=1,2,...$
they average out. If $P>T_{\rm obs}$, as is the case for the
insidious 18.6-year lunar tide acting on the node and the perigee,
the $K_1,\ l=3,\ m=1,\ p=1,\ q=-1$ ocean tide, whose period
amounts to 5.07 years, acting on the perigee of LAGEOS II, and a
harmonic constituent due to the direct solar radiation pressure,
whose period amounts to 11.6 years, acting on the perigee of
LAGEOS II as well\footnote{Notice that the period of the
18.6--year tide depends only on the luni--solar variables, while
the periods of the $K_1$ tide and of the direct solar radiation
pressure harmonic depend on the orbital geometry of the
satellite.}, they may alias the relativistic trend acting as
superimposed linear signals. However, the 18.6-year tide is an
even zonal perturbation which can be canceled out by designing
suitably the observable, as done with \rfr{uno}. The impact of the
other two mismodelled harmonics is reduced by the small value of
the coefficient $c_2$ with which the perigee of LAGEOS II enters
the combination of \rfr{uno}
\item Secular aliasing
trends of gravitational (static even zonal harmonics of the
geopotential [{\it Kaula}, 1966]) and non-gravitational (some tiny
thermal effects [{\it Lucchesi}, 2001b]) origin. The impact of the
mismodelled even zonal harmonics is far the most relevant in the
error budget. Indeed, the systematic error $\delta\mu_{\rm LT}$
induced by them cannot be suitably handled or removed from the
signal. One can only assess as more reliably as possible its
impact on the measurement of $\mu_{\rm LT}$.

\Rfr{uno} cancels out the first two even zonal harmonics: the
effect of the remaining higher degree mismodelled harmonics
$\delta J_6,\ \delta J_8,...$ amounts to 12.9$\%$ according to the
covariance matrix of EGM96 Earth gravity model [{\it Lemoine et
al.}, 1998] up to degree $l=20$\footnote{In regard to this point,
it should be mentioned that, at present, there is no full
consensus in the scientific community on the reliability of such
estimate which might be rather optimistic [{\it Ries et al.},
1998]. Indeed, it would not be entirely correct to automatically
extend the validity of the covariance matrix of EGM96, which is
based on a multi--year average that spans the 1970, 1980 and early
1990 decades, to any particular time span like that, e.g., of the
LAGEOS--LAGEOS II analysis which extends from the middle to the
end of the 1990 decade. Indeed, there would not be assurance that
the errors in the even zonal harmonics of the geopotential during
the time of the LAGEOS--LAGEOS II experiment remained correlated
exactly as in the EGM96 covariance matrix, in view of the various
secular, seasonal and stochastic variations that we know occur in
the terrestrial gravitational field and that have been neglected
in the EGM96 solution.}.
\end{itemize}
%--------------------------------------------------------------------------------
\section{The search for alternative combinations}
A possible strategy for improving the accuracy of the present--day
Lense--Thirring LAGEOS--LAGEOS II experiment consists of suitable
combinations of the orbital residuals $\delta\dot\Omega$ and
$\delta\dot\omega$ of the rates of the nodes and the perigees of
different SLR satellites. Such combinations can be written in the
form \eqi\sum_{i=1}^N c_i f_i=X_{\rm GR}\mu_{\rm
GR},\lb{combi}\eqf in which the coefficients $c_i$ are, in
general, suitably built up with the orbital parameters of the
satellites entering the combinations, the $f_i$ are the residuals
$\delta\dot\Omega,\ \delta\dot\omega$ of the rates of the nodes
and the perigees of the satellites entering the combination,
$X_{\rm GR}$ is the slope, in mas/y, of the general relativistic
trend of interest and $\mu_{\rm GR}$ is the solve--for parameter,
to be determined by means of usual least--square procedures, which
accounts for the general relativistic effect. For example, in the
case of the Lense--Thirring LAGEOS--LAGEOS II experiment [{\it
Ciufolini}, 1996] $X_{\rm LT}=60.2$ mas/y. More precisely, the
combinations of \rfr{combi} are obtained in the following way. The
equations for the residuals of the rates of the $N$ chosen orbital
elements are written down, so to obtain a non homogeneous
algebraic linear system of $N$ equations in $N$ unknowns. They are
$\mu_{\rm GR}$ and the first $N-1$ mismodelled spherical harmonics
coefficients $\delta J_l$ in terms of which the residual rates are
expressed [{\it Iorio}, 2002c]. The coefficients $c_i$ and,
consequently, $X_{\rm GR}$ are obtained by solving for $\mu_{\rm
GR}$ the system of equations. So, the coefficients $c_i$ are
calculated in order to cancel out the contributions of the first
$N-1$ even zonal mismodelled harmonics which represent the major
source of uncertainty in the Lense--Thirring precessions. The
coefficients $c_i$ can be either constant \footnote{In general,
the coefficient of the first orbital element entering a given
combination is equal to 1, as for the combinations in [{\it
Ciufolini}, 1996; {\it Iorio}, 2002b; {\it Iorio et al.}, 2002a;
2002b]. } or depend on the orbital elements of the satellites
entering the combinations.

Now we expose how to calculate the systematic error due to the
mismodelled even zonal harmonics of the geopotential for the
combinations involving the residuals of the nodes and the perigees
of various satellites.

In general, if we have an observable $q$ which is a function
$q=q(x_j)$, $j=1,2...M$ of $M$ correlated parameters $x_j$ the
error in it is given by
 \eqi \delta
q=\left[\sum_{j=1}^M\left(\derp{q}{x_j}\right)^{2}\sigma_j^2+2\sum_{h\neq
k
=1}^M\left(\derp{q}{x_h}\right)\left(\derp{q}{x_k}\right)\sigma^{2}_{hk}\right]^{\frac{1}{2}}\lb{app1}\eqf
in which $\sigma^{2}_{j}\equiv C_{jj}$ and $\sigma^{2}_{hk}\equiv
C_{hk}$ where $\{C_{hk}\}$ is the square matrix of covariance of
the parameters $x_j$.

In our case the observable $q$ is any residuals' combination \eqi
q=\sum_{i=1}^{N}c_i f_i(x_j),\ j=1,2...10,\eqf where $x_j,\
j=1,2...10$ are the even zonal geopotential's coefficients $J_2,\
J_4...J_{20}$.  Since \eqi \derp{q}{x_j}=\sum_{i=1}^N c_i
\derp{f_i}{x_j},\ j=1,2...10\lb{app2},\eqf by putting \rfr{app2}
in \rfr{app1} one obtains, in mas/y \eqi \delta
q=\left[\sum_{j=1}^{10}\left(\sum_{i=1}^N c_i
\derp{f_i}{x_j}\right)^{2}\sigma_j^2+2\sum_{h\neq k
=1}^{10}\left(\sum_{i=1}^N c_i
\derp{f_i}{x_h}\right)\left(\sum_{i=1}^N c_i
\derp{f_i}{x_k}\right)\sigma^{2}_{hk}\right]^{\frac{1}{2}}.\lb{app3}\eqf
The percent error, for a given general relativistic trend and for
a given combination, is obtained by taking the ratio of \rfr{app3}
to the slope in mas yr$^{-1}$ of the general relativistic trend
for the residual combination considered.

The validity of \rfr{app3} has been checked by calculating with it
and the covariance matrix of the EGM96 gravity model up to degree
$l=20$ the systematic error due to the even zonal harmonics of the
geopotential of the Lense--Thirring LAGEOS--LAGEOS II experiment;
indeed the result \eqi\delta\mu_{\rm LT}=13\%\ \mu_{\rm LT}\eqf
claimed in [{\it Ciufolini et al.}, 1998] has been obtained
again\footnote{It may be interesting to notice that if, with a
more conservative approach, we consider only the diagonal part of
the EGM96 covariance matrix, up to degree $l=20$, the systematic
error due to the mismodelled even zonal harmonics of the
geopotential for the Lense--Thirring LAGEOS--LAGEOS II experiment
amounts to 46.5$\%$ [{\it Iorio,} 2002c].}. For the systematic
errors due to the even zonal harmonics of the geopotential of
alternative proposed gravitomagnetic and gravitoelectric
experiments, according to EGM96, see [{\it Iorio}, 2002b; {\it
Iorio et al.}, 2002a; 2002b]. It is worth noticing that, since the
orbits of the LAGEOS satellites are almost insensitive to the
geopotential's terms of degree higher than $l=20$, the estimates
based on the covariance matrix of the EGM96 gravity model up to
degree $l=20$ should be considered rather reliable although the
higher degree terms of EGM96 might be determined with a low
accuracy.

It should also be pointed out that the evaluations of the
systematic error due to geopotential based on the approach of the
combined orbital residuals should be free from some uncertainties
due to possible secular, seasonal and stochastic effects. Indeed,
even putting aside the fact that most of the secular and seasonal
variations of the geopotential are concentrated just in the first
even zonal harmonics, if we cancel out as many even zonal
harmonics as possible, the uncertainties in the evaluation of the
error based on the remaining correlated even zonal harmonics of
higher degree should be greatly reduced irrespectively of the
chosen time span.

A very important point to stress is that the forthcoming new data
on the Earth's gravitational field by CHAMP, which has been
launched in July 2000, and GRACE, which has been launched in March
2002, will have a great impact on the reduction of the systematic
error due to the mismodelled part of geopotential and on the
reliability of the estimates based on it.
%-----------------------------------------------------------------------------
\subsection{The combinations without the perigee of LAGEOS II} In
regard to alternative combinations, we first explore the
possibility of reducing $\delta\mu_{\rm LT}$ by adopting different
residual combinations obtained by substituting the perigee of
LAGEOS II\footnote{The orbits of the other SLR satellites are far
less elliptical than that of LAGEOS II (except for Starlette), so
that their perigees are not good observables. It turns out that
the inclusion of the perigee of Starlette would not yield any
improvements.} with one or more nodes of Ajisai, Starlette,
Stella, WESTSPAC­-1, ETALON­-1 and ETALON­-2. The advantages of
using only the nodes would be
\begin{itemize}
\item No odd zonal geopotential harmonics
\item No odd degree ocean tidal perturbations
\item Non­-gravitational perturbations of smaller magnitude and
fairly well modelled \item Atmospheric drag negligible not only
for the LAGEOS and the ETALON satellites, but also for Starlette
and Ajisai orbiting at lower altitudes
\item Smaller rms on the experimental residuals
\end{itemize}

The exclusion of the perigee of LAGEOS II leads to two classes of
residual combinations.
\begin{itemize}
\item Combinations without
the ETALON satellites. They should be discarded because the error
due to the uncancelled higher degree even zonal harmonics of the
geopotential, according to EGM96, is itself far larger than
20$\%$, which is the estimated total error in the LAGEOS--LAGEOS
II Lense--Thirring measurement. It is so because of the fact that
the satellites orbiting at lower altitudes than LAGEOS and LAGEOS
II are more sensitive to the higher degree even zonal harmonics of
the geopotential.
\item Combinations with the ETALON satellites. Due to their higher
altitude with respect to the LAGEOS satellites $\delta\mu_{\rm
LT}$ falls, in some cases, below 1$\%$. However, they should not
be considered because the coefficients of the nodes of the ETALON
satellites are, in these cases, of the order of $10^ 3$ so that
they greatly amplify every perturbations acting on such elements.
This is particularly true for the solid Earth tesseral $K_1$ tide.
Indeed, it turns out that its $l=2,\ m=1,\ p=1,\ q=0$ constituent
induces on ETALON--1 a perturbation with a nominal amplitude of
­1,216.57 mas and period of ­10,880.8 days while for ETALON-­2 we
have a nominal amplitude of ­1,269.35 mas and a period of
­11,130.1 days.  Even if the solid Earth tides are known at a
0.5$\%$ level [{\it Iorio}, 2001], the effect of the too large
weighing coefficients and the very long periods of such tidal
perturbations would induce insidious mismodelled linear trends
over reasonable observational time spans of few years. However, it
must pointed out that a practical useful use of the SLR data of
the ETALON satellites would be problematic mainly because the
Russian satellites are, at present, very poorly tracked. Moreover,
they wobble and their center of mass is not well defined. Last but
not least, the accuracy with which they were constructed and the
pre­launch ground controls are not comparable to those of LAGEOS
satellites.
\end{itemize}
In conclusion, even if very appealing, the possibility of using
residual combinations involving only the nodes of the existing
geodetic SLR satellites should be rejected.
%-----------------------------------------------------------------------------------------------
\subsection{The combinations with the perigee of LAGEOS II} It
turns out that it is possible to obtain admissible results only by
involving the perigee of LAGEOS II.
\begin{itemize}
\item Combinations with ETALON­-1. We obtained seventeen
combinations which, in principle, could be considered because
$\delta\mu_{\rm LT}$ $<$ 20$\%$ for them. However, it turns out
that the combinations including the perigee of LAGEOS II and the
node of ETALON--1 cannot represent a genuine improvement with
respect to the present day LAGEOS--LAGEOS II Lense--Thirring
experiment because $\delta\mu_{\rm LT}\geq 13\%$ for them. The
systematic errors due to the time--dependent non--gravitational
perturbations would make the total error budget unfavorable with
respect to the current experiment.
\item Combinations without ETALON­-1. We obtained eight combinations
involving only the LAGEOS satellites and the lower altitude
satellites. It turns out that the combination
\eqi\delta\dot\Omega^{\rm I}+c_1\delta\dot\Omega^{\rm
II}+c_2\delta\dot\Omega^{\rm Aj}+c_3\delta\dot\omega^{\rm II}\sim
\mu_{\rm LT}61.2\lb{due},\eqf with $c_1=0.443,\ c_2=-0.0275,\
c_3=-0.341$ is the only one that could lead to a real, slight
improvement of the error budget\footnote{It may be interesting to
notice that if, with a more conservative approach, we consider the
diagonal part only of the EGM96 covariance matrix, up to degree
$l=20$, the systematic error due to the mismodelled even zonal
harmonics of the geopotential for such alternative combination
amounts to 64.2$\%$ [{\it Iorio,} 2002c].} [{\it Iorio}, 2002b].
Indeed
\begin{itemize}
\item
The systematic error due to the even zonal harmonics amount to
$\delta\mu_{\rm LT}\sim 10.7\%$ since \rfr{due} cancels out the
effect of $\delta J_6$ as well
\item The estimates relative to the periodic time--varying
perturbations for \rfr{uno} would be valid also in this case
because the coefficients of the node and the perigee of LAGEOS II
are almost the same \item The additional perturbations induced by
Ajisai are damped since the coefficient of its node amounts only
to -0.0275
\item It turns out that the tidal perturbations on the node of Ajisai
have periods not longer than few years and the non­-gravitational
perturbations are fairly well studied [{\it Sengoku et al.}, 1995;
1997]
\item Ajisai, contrary to the ETALON
satellites, is a well known and accurately tracked geodetic
satellite \end{itemize}
\end{itemize}
%Moreover, within few years the new data on
%the geopotential from the CHAMP and GRACE missions should be
%available; they should improve our knowledge, among other things,
%of the spherical harmonics coefficients of the geopotential.
%Since
%Ajisai is more sensible than LAGEOS to them it could be expected
%that $\delta\mu_{\rm LT}$ of \rfr{due} will diminish more than
%$\delta\mu_{\rm LT}$ of \rfr{uno}.
%---------------------------------------------------------------------
\section{Conclusions}
The combination of \rfr{uno} plays a really important role in the
context of the efforts aimed to the detection of the general
relativistic Lense-Thirring drag in the gravitational field of the
Earth. Indeed, at present, it represents a relatively fast, cheap,
reliable and almost unique way for measuring this effect in the
terrestrial space environment by means of the analysis of the
orbital data of the existing laser--tracked satellites. This
should also justify and encourage the efforts aimed to improve the
error budget and the knowledge of the aliasing classical forces
affecting such measurement. However, according to the covariance
matrix of the EGM96 model up to degree $l=20$, the alternative
combination of \rfr{due} might enforce and slightly improve the
experimental accuracy of the measurement.

At the present--day level of knowledge of the perturbing forces of
the terrestrial space environment, only the not yet approved
LAGEOS--LARES mission, whose originally proposed configuration has
been recently modified [{\it Iorio et al.}, 2002b], could yield a
notable, genuine improvement in measuring the Lense--Thirring
effect with SLR.

Moreover, within few years the new data on the geopotential from
the CHAMP and GRACE missions should be available. They should
improve our knowledge, among other things, of the spherical
harmonics coefficients of the geopotential, and, consequently, the
systematic gravitational errors in all the proposed
Lense--Thirring measurements should reduce as well.
%-----------------------------------------

\end{document}